# Enhanced Electron Extraction in Co-Doped TiO$_2$ Quantified by Drift-Diffusion Simulation for Stable CsPbI$_3$ Solar Cells

Thomas W. Gries[1,2], Davide Regaldo[3,4,5], Hans Köbler[1], Noor Titan Putri Hartono[1], Gennaro V. Sannino[6], Emilio Gutierrez Partida[7], Roberto Félix[1], Elif Hüsam[1], Ahmed Saleh[1], Regan G. Wilks[1], Zafar Iqbal[1], Zahra Loghman Nia[1], Martin Stolterfoht[7,8], Marcus Bär[1,9,10], Stefan A. Weber[11], Paola Delli Veneri[6], Philip Schulz[3,12], Jean-Baptiste Puel[3,13], Jean-Paul Kleider[3,4,5], Eva Unger[1], Qiong Wang*,[1], Artem Musiienko*,[1] and Antonio Abate*,[1,2]

[1]Helmholtz-Zentrum Berlin für Materialien und Energie (HZB), Berlin, Germany.
[2]Department of Chemistry, University of Bielefeld, Bielefeld, Germany.
[3]Institut Photovoltaïque d'Ile-de-France (IPVF), Palaiseau, France.
[4]Université Paris-Saclay, CentraleSupélec, CNRS, Laboratoire de Génie Electrique et Electronique de Paris, Gif-sur-Yvette, France.
[5]Sorbonne Université, CNRS, Laboratoire de Génie Electrique et Electronique de Paris, 75252, Paris, France.
[6]Portici Research Center, Italian National Agency for New Technologies, Energy and Sustainable Economic Development (ENEA), Portici, Naples, Italy.
[7]Institute for Physics and Astronomy, University of Potsdam, Potsdam, Germany.
[8]Electronic Engineering Department, The Chinese University of Hong Kong, Shatin, N.T., Hong Kong.
[9]Department of Physical Chemistry II, Friedrich-Alexander-University Erlangen-Nürnberg, Erlangen, Germany.
[10]Department of X-Ray Spectroscopy at Interfaces of Thin Films, Helmholtz-Institute Erlangen-Nürnberg for Renewable Energy, Berlin, Germany.
[11]Institute for Photovoltaics, University of Stuttgart, Stuttgart, Germany.
[12]Ecole Polytechnique, IPVF, UMR 9006, CNRS, Palaiseau, France.
[13]EDF R&D, Palaiseau, France.

# Abstract

Solar cells based on inorganic perovskite $CsPbI_3$ are promising candidates to resolve the challenge of operational stability in the field of perovskite photovoltaics. For stable operation, however, it is crucial to thoroughly understand the extractive and recombinative processes occurring at the interfaces of perovskite and the charge-selective layers. In this study, we focus on the electronic properties of (doped) $TiO_2$ as an electron-selective contact. We show via KPFM that co-doping of $TiO_2$ with Nb(V) and Sn(IV) reduces the material's work function by 270 meV, giving it stronger n-type characteristics compared to Nb(V) mono-doped $TiO_2$. The modified electronic alignment with $CsPbI_3$ translates to enhanced electron extraction, as demonstrated with steady-state photoluminescence spectroscopy, transient photoluminescence and transient surface photovoltage in triad. Importantly, we extract crucial parameters, such as the concentration of extracted electrons and the interface hole recombination velocity, from the SPV transients via 2D drift-diffusion simulations. When implementing the co-doped $TiO_2$ into full n-*i*-p solar cells, the operational stability under continuous AM1.5G illumination is enhanced from 970 h to 25'000 h of projected $T_{S80}$ lifetime. This study provides fundamental understanding of interfacial charge extraction and its correlation with operational stability of perovskite solar cells, which can be transferred to other charge-selective contacts.

# Introduction

Stable operation is one of the key factors for the successful market introduction of perovskite solar cells (PSCs). To achieve stable operation, photoinduced degradation processes in the perovskite light-absorbing layer must be minimized. Organic cations, such as methylammonium (MA$^+$), are prone to intrinsic destabilization of the perovskite absorber due to their volatile nature.[1-4] From this perspective, inorganic perovskites, such as CsPbI$_3$ (bandgap: $E_g$ = 1.7 eV), are attractive candidates to achieve stable PSCs.[1][5][6] Recently, Zhao et al. have demonstrated impressively stable solar cells based on CsPbI$_3$ with no detectable degradation after 3500 h at 35 °C.[5]

In the field of inorganic perovskite CsPbI$_3$ absorbers, most research efforts so far have targeted the crystallization dynamics via additive engineering, lowering the phase transition temperature to stabilize the metastable black β- and γ-phases (tetragonal and orthorhombic) at room temperature, as well as interfacial defect passivation.[7-11] Those strategies have led to efficiencies exceeding 21% with defect passivation strategies mainly contributing to an increase in open-circuit voltage ($V_{OC}$).[11][12]

The wide-bandgap semiconductor titanium dioxide (TiO$_2$) in its anatase crystal phase is generally suitable as an electron-selective contact (ESC) in PSCs due to its wide $E_g$ of 3.2 eV, and its electron affinity around -4.0 eV with respect to the vacuum level ($E_{vac}$) reported in the literature.[13-16] In fact, best-performing n-*i*-p structured PSCs based on inorganic CsPbI$_3$ use either TiO$_2$ or SnO$_2$ as ESC and spiro-OMeTAD as hole-selective contact (HSC).[7][8][10][12][17][18] However, defective TiO$_2$ surfaces, especially surface oxygen vacancies and Ti(III)-derived states serve as non-radiative recombination centers and are suspected to serve as oxidation sites for iodide to elemental iodine under irradiation, potentially leading to degradation of the adjacent perovskite.[19][20] A minimization of surface defects is therefore needed to improve charge extraction and the stability of the resulting solar cells. Moreover, the low electron conductivity of pristine TiO$_2$ results in an increased series resistance and, therefore, reduces the fill factor (FF).[21]

In our study, we introduce a co-doping strategy for TiO$_2$, which reduces the density of interfacial defects at the ESC/CsPbI$_3$ interface. Aliovalent doping of anatase TiO$_2$ with Nb(V) is an established procedure for enhanced n-type characteristics of the material.[22][23] Nb(V) enhances the conductivity of the material by releasing additional electrons into the conduction band (n-type effect).[22][24] The concentration-dependent decrease in resistivity reaches a plateau for Nb(V) above 3%, with values as low as 2.3$10^{-4}$ $\Omega$cm$^{-2}$ at 300 K.[22] Moreover, Nb(V) inhibits the transition to undesirable rutile TiO$_2$.[25] Nb(V)-doped TiO$_2$ has first been applied in dye-sensitized solar cells in 2010 and in perovskite solar cells in 2015, where the improved photovoltaic performance at a doping level of 0.5% Nb(V) was ascribed to suppressed Ti(III) defect creation.[26][27] However, the achievable upward shift of the Fermi level in anatase TiO$_2$ by using Nb(V) is limited since the Fermi level decreases again beyond the maximum value, possibly due to localized Ti 3d$^1$ states in highly Nb(V)-doped TiO$_2$.[26][28][29] In this study, we add isovalent Sn(IV) as a co-dopant to Nb(V)-doped TiO$_2$ to further increase the n-type character of TiO$_2$. By combining both dopants, we surpass the limit of the single dopant, and achieve dopant-concentration-dependent upwards tunability of the Fermi-level position in TiO$_2$.

To unveil the effect of co-doping on the fundamental charge extraction and recombination mechanisms, we applied time-resolved surface photovoltage (trSPV) and drift-diffusion (DD) modeling. Qualitative assessment of recorded transient curves is frequently done in the photovoltaic community.[30-32] However, quantitative elucidation of extraction and recombination velocities requires precise knowledge of the material energetics and advanced curve fitting models. So far, the minimalistic kinetic model developed by Levine and Musiienko is mostly used.[8][30][33] The minimalistic model lacks incorporation of drift and diffusion phenomena, as well as consideration of the alignment of interfaces and the spatial distribution of charge carriers. These drawbacks restrict our ability to acquire comprehensive insights into interface characteristics through time-resolved measurements. Here, we developed an approach of fitting SPV transients with the DD model, including drift and diffusion phenomena. With the help of the DD model, we reveal fundamental properties of electron extraction and recombination. We show that $TiO_2$ co-doped with 0.5% Nb(V) and 0.1% Sn(IV) exhibits strongly reduced recombination of electrons with hole minority carriers. This reduction is expressed in an interfacial hole recombination velocity of 0.098 cm, representing a reduction by two orders of magnitude compared to 17.0 cm for $TiO_2$ mono-doped with 0.5% Nb(V). Consequently, the concentration of extracted electrons is enhanced from $3.7 10^{10}$ cm$^{-2}$ to $4.8 10^{10}$ cm$^{-2}$, respectively, as detailed below.

The lower interfacial hole recombination velocity and increased number of extracted electrons for co-doped $TiO_2$ is reflected in an improved PSC performance. PCE is enhanced by 1.0% absolute on average, mainly originating from a statistical improvement in FF by 2% to 80.8%. Further, $V_{OC}$ is slightly improved by 10 mV to 1.18 V. The lowered extraction barrier in n-type co-doped $TiO_2$ and the reduced hole recombination velocities are responsible for both FF and $V_{OC}$ improvement.[34] Most importantly, the co-doping approach translates to a drastic stabilization in MPP tracking under continuous AM1.5G illumination, where no significant degradation is detected within the first 300 h. As a result, the projected $T_{S80}$ lifetime improves from 970 h in mono-doped $TiO_2$ to 25'000 h in co-doped $TiO_2$

## Results and Discussion

### Energetic Manipulation of the Electron Selective Contact

Doped $TiO_2$ ESCs were deposited using the spray pyrolysis method, providing an oxidative environment via the $O_2$ carrier gas. The process yielded compact layers of 20 nm thickness, as confirmed by variable-angle spectroscopic ellipsometry (VASE, **Fig. S1**). To quantify the effect of dopant addition on the Fermi level position of the as-deposited TiO$_2$, we measured the surface potential via Kelvin-probe force microscopy (KPFM) in the dark. The details about the measurement conditions and data treatment are given in **Sec. S3.3** in SI. Fig. 1**B** shows that with Nb(V)-doping alone, a reduction of the work function $\phi$ (WF) by 270 meV could be achieved at the optimum concentration of 0.5% Nb(V). The same dopant concentration was confirmed as optimum in other studies, although WF changes have not been characterized.[26][27] Furthermore, upon the addition of 0.1% of co-dopant Sn(IV), we observed an even lower WF by 80 meV compared to the mono-doped champion. The full optimization series of co-doped TiO$_2$ was based on photovoltaic performance and is shown in **Fig. S11** and **Fig. S12**. Assuming vacuum level alignment, the reduction of the WF can be interpreted as a raise of the Fermi level. It is accompanied by a slight increase in $E_g$, as estimated from UV/Vis spectra via $(\alpha h\nu)^{1/2}$ vs. $h\nu$ plots (Tauc analysis for indirect semiconductors, **Fig. S2**). Bandgap widening in the same range of 40 meV has been observed for Sn(IV)-doped TiO$_2$ and has been attributed to the mixing of Sn 5s and Ti 3d states at the conduction band minimum (CBM).[35][36]

In both cases, the WF increases again beyond the optimum concentrations of 0.5% Nb(V) and 0.1% Sn(IV), respectively. This increase can be associated with the creation of intragap defect states, such as the Ti 3d$^1$ state for Nb(V) doped anatase mentioned above or, in case of Sn(IV) doping, a facilitated phase transition to rutile TiO$_2$ which possesses a deeper $E_F$.[35] The hypothesis of the partial transition to rutile ($E_g = 3.0$ eV[37]) is supported by a reversal of bandgap widening beyond the optimum Sn(IV) concentration (**Fig. S2**).[36]

Beyond quantification of the average shift of the WF, effects such as phase segregation and transformation can be indicated in the full width at half maximum (FWHM) of the contact potential difference (CPD) histograms obtained on (5 x 5) μm$^2$ KPFM scans (Fig. 1**B**). Compared to the reference line width of 25 meV obtained on non-doped TiO$_2$, only the sample co-doped with 2.0% Sn(IV) showed significant broadening. This broadening further indicates the phase transition to rutile TiO$_2$ promoted by Sn(IV) as suggested above. In all other samples, the absence of line broadening indicates a homogenous phase formation without segregation phenomena.

Fig. 1**C** shows the energetic positioning of the mono-doped TiO$_2$ (green) and the co-doped TiO$_2$ (blue) in relation to CsPbI$_3$ (brown). The WF of 3.49 eV for CsPbI$_3$ measured with KPFM in the dark is comparable to values measured with UPS reported elsewhere (for details see **Sec. S3.3** in the SI).[8] $E_F - E_{VBM}$ distances are determined via hard X-ray photoelectron spectroscopy (HAXPES), as described in **Fig. SX** in the Supporting Information (SI), while values for $E_{CBM}$ can only be estimated from addition of $E_g$ to $E_{VBM}$. By co-doping TiO$_2$, the energetic misalignment to CsPbI$_3$ is mediated. In addition, the conductivity of TiO$_2$ is enhanced due to the shift of $E_F$ towards $E_{CBM}$.

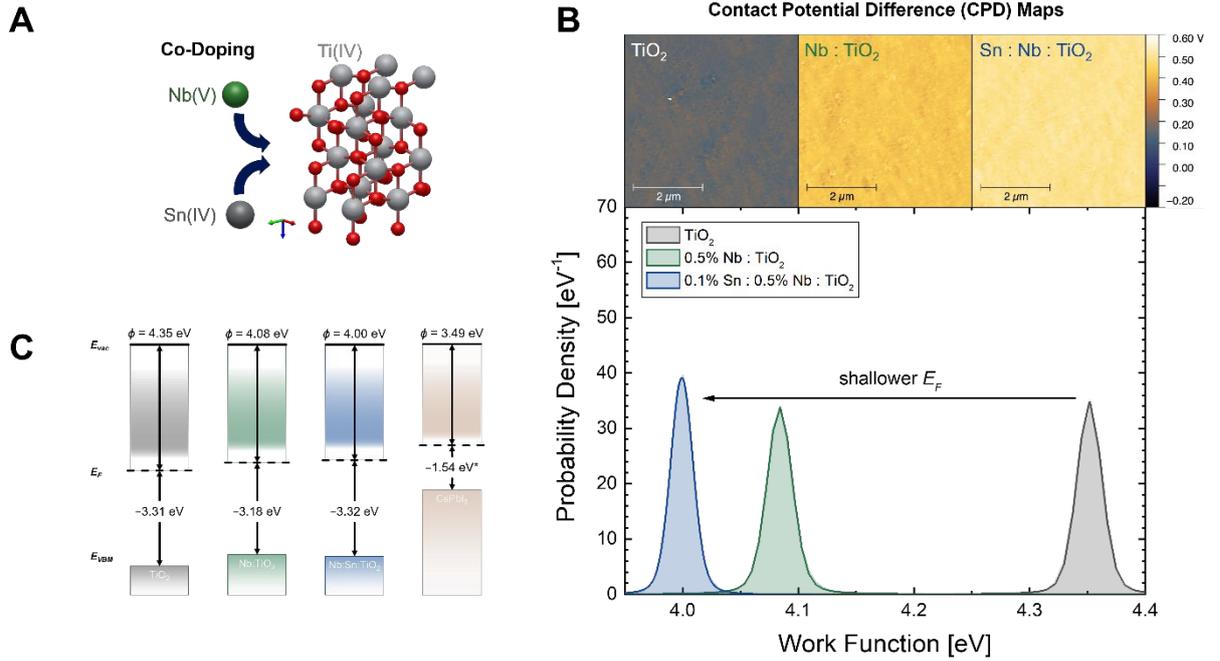

**Fig. 1. The WF of TiO$_2$ can be tuned via co-doping (A)** Nb(V) and Sn(IV) are added to TiO$_2$, represented as supercell of anatase (space group 141, $I4_1/amd$). **(B)** The maximum reduction of the WF of Nb(V) mono-doped TiO$_2$ can be overcome by co-doping with Sn(IV). All curves are extracted from (5 x 5) µm$^2$ KPFM scans in argon atmosphere, minimizing the influence of water adsorption on the measured WF values. **(C)** Schematic energy level representation of the CsPbI$_3$ (depicted in brown) in relation to non-doped (gray), mono-doped (green), and co-doped (blue) TiO$_2$, where the co-doped sample shows the highest degree of n-type doping, explaining the lowest WF. The scheme assumes the separate materials with aligned vacuum level in dark conditions. The value marked with an asterisk was measured in a previous study.[8]

The chemical structure at the surface of the (co-)doped sample series was further investigated via synchrotron-based HAXPES, allowing for higher resolution of spectral line shapes than laboratory-based X-ray photoelectron spectroscopy instruments. Of special interest was the detection of the low concentration dopants and their impact on the Ti chemical environment of the samples. For this purpose, thin films of $TiO_2$ (i.e., 20 nm nominal thickness) with corresponding (co-)doping levels were deposited on fluorine-doped tin oxide (FTO) coated glass substrates (further sample preparation details are described in **Sec. S2.3** in SI). The HAXPES survey spectra of the investigated samples are shown in **Fig. S7** in SI. Inspecting the survey spectra reveals that signal detected from the samples originates predominantly from the $TiO_2$ layers; however, a minor (yet significant) and varying fraction of the signal derives from the FTO substrate, indicating an incomplete coverage. As the FTO contains Sn(IV), we refrain from assessing the chemical environment and quantity of the Sn(IV) dopant, especially considering the low Sn(IV) (co-)doping levels (i.e., 0.1% and 2%) used in this study. Fig. 2 shows HAXPES detail spectra of the Ti 2p (**A**) and Nb 3d (**B**) energy regions of the investigated samples series, measured with 2 keV excitation, including curve fit results (for further details on the curve fit analysis of measured spectra, see **Sec. S1**). As shown in Fig. 2**A**, the binding energy (BE) position of the Ti $2p_{3/2}$ line is found at $(459.3 \pm 0.1)$ eV, consistent with reports for $TiO_2$ in literature.[38] Moreover, no discernible signal is detected at BE ranges reported for Ti(III) (e.g., $Ti_2O_3$) or Ti(II) (e.g., TiO) in literature,[38][39] as denoted by the gray-filled areas in the figure. The highly similar line shape of the Ti 2p spectra for all investigated samples, with only one set of doublet peaks needed to obtain good fit results, indicates a homogenous Ti chemical environment throughout the sample series, comprised by one Ti chemical species (i.e., $TiO_2$). Likewise, the Nb 3d spectra of all Nb-doped samples, shown in Fig. 2**B**, can be modeled with one set of doublet peaks, indicating a common Nb chemical environment. The Nb $3d_{5/2}$ line is located at a BE value of $(207.8 \pm 0.1)$ eV, matching reports of Nb(V) (e.g., $Nb_2O_5$) in literature.[38] Based on the intensities of the Ti 2p and Nb 3d measurements, the [Nb]:[Ti] composition ratio of the samples can be computed (for further details on HAXPES-derived quantification, see **Sec. S3.4** in SI).

The results of this quantification are shown in Fig. 2**C**, which are in excellent agreement with the nominal Nb(V)-doping concentration.

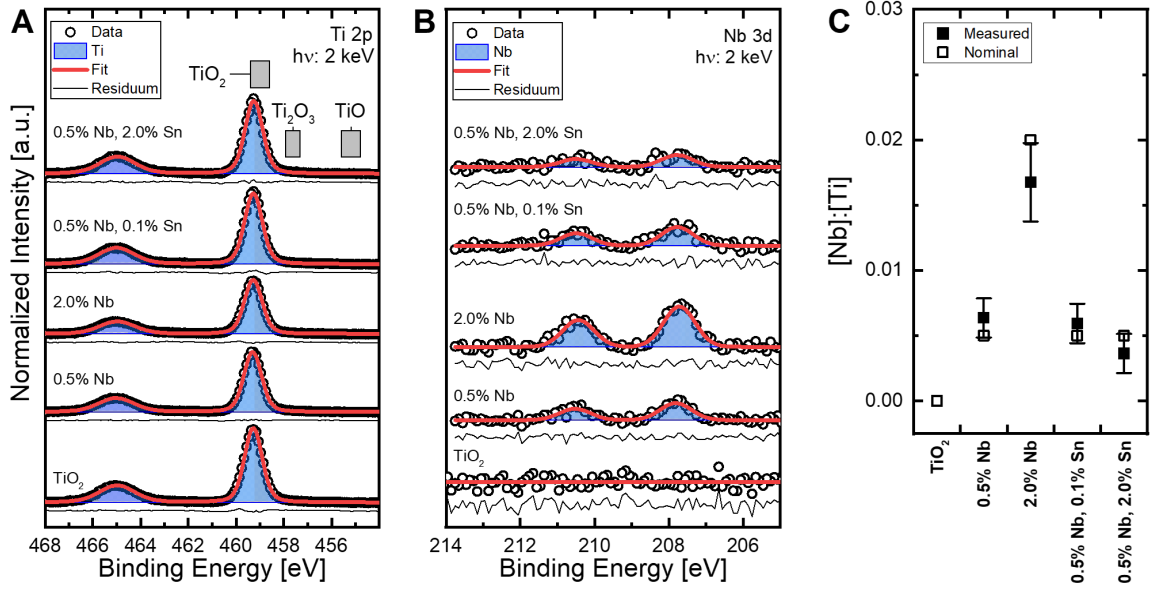

**Fig. 2. Nb(V) dopant levels agree with nominal values.** HAXPES detail spectra of the **(A)** Ti 2p and **(B)** Nb 3d energy regions of the differently (co-)doped $TiO_2$ samples. The spectra were measured using a 2 keV excitation and normalized to background intensity, with vertical offsets added for clarity. Curve fit results are included. The gray-filled areas in (A) are binding energy (BE) ranges reported in the literature for Ti-based reference compounds.[38][39] **(C)** [Nb]:[Ti] surface composition of variously treated $TiO_2$ films, determined by the HAXPES measurements shown in (A) and (B).

The KPFM maps show lateral homogeneity, suggesting that the derived WF values are not influenced by the coverage issues observed using HAXPES. The KPFM results indicate that the position of the Fermi level of the $TiO_2$ ESC can be raised by the addition of 0.5% Nb(V) and 0.1% Sn(IV), while HAXPES Ti 2p results show no discernible Ti(III) contribution for these samples (Fig. 2**A**). Therefore, doping does not affect the density of Ti(III) defect states. The associated increase in conductivity reduces the series resistance and should mainly lead to an increase in the fill factor of respective solar cells, as discussed below. Next, we focus on the interfacial charge carrier dynamics when the doped $TiO_2$ systems are put into contact with $CsPbI_3$ perovskite.

## Interface Characterization

To fully characterize the impact of doping on non-radiative recombination and charge extraction dynamics, we measured absolute steady-state photoluminescence (ssPL), transient photoluminescence (trPL) and transient surface photovoltage (trSPV) on photovoltaic half-cells with (doped) $TiO_2$ and $CsPbI_3$ (Fig. 3).

Since a reversible shift in $E_F$ is observed with a maximum at 0.1% Sn(IV) co-dopant compared to the sample "over-doped" at 2.0% Sn(IV), we investigated those samples with ssPL. Increasing the co-dopant level of Sn(IV) to 2.0% leads to steady decrease of the PL peak amplitude (**Fig. S8B**) to 70% of the Sn(IV)-free intensity, indicating decreased radiative recombination with increased Sn(IV) concentration. Steady-state measurements alone cannot provide detailed insights into the interfacial charge-carrier dynamics. We therefore employed time-resolved spectroscopic methods for further characterization.

We analyzed the transient decay of the PL signal obtained in trPL (Fig. 3**B**) via biexponential fitting. The carrier lifetime associated with the fast component of the biexponential decay ($\tau_{fast}$) decreased from 5 ns to 4 ns upon addition of 0.1% Sn(IV) co-dopant compared to the 0.5% Nb(V) mono-doped $TiO_2$ sample, while the slow component ($\tau_{slow}$) decreased from 32 ns to 22 ns. Since the perovskite preparation and crystallization are the same in all cases, the bulk lifetime resulting from crystal quality is not expected to change, also consistent with the unchanged FWHM in the ssPL signal.

Evidently, in a bilayer system the trPL decay is influenced not only by non-radiative recombination at defect sites and interface recombination but, importantly, also by charge carriers extracted to the ESC.[40] Specifically, electron extraction leads to a quick decrease of the photoluminescence signal at short timescales, dominating $\tau_{fast}$. Since electron extraction to $TiO_2$, in turn, is dependent on the extraction rate, the interface recombination rate, and conduction band offset to CsPbI$_3$. Once a maximum of electrons is accumulated in TiO$_2$, the PL signal decay will slow down, and back-transfer to CsPbI$_3$ or recombination with holes in CsPbI$_3$ is favored (s. Fig. 4**E**). Therefore, the extraction process also becomes crucial when discussing $\tau_{slow}$, especially at medium laser fluences of 30 nJcm$^{-2}$ as used in this study.[40]

Sn(IV) addition shortens both $\tau_{fast}$ and $\tau_{slow}$. We suggest that those shortened lifetimes are a direct consequence of the improved energetic alignment between co-doped TiO$_2$ and CsPbI$_3$, as demonstrated via KPFM and HAXPES (s. Fig. 1**C**). Closer matching of the WFs of TiO$_2$ and CsPbI$_3$ leads to faster extraction to the ESC, responsible for the lower $\tau_{fast}$ component. Accordingly, also electron back transfer from co-doped TiO$_2$ to CsPbI$_3$ is faster, leading to comparatively higher carrier concentrations at longer timescales and, therefore, reducing the $\tau_{slow}$ component. Since it is difficult to disentangle electron back transfer and interface recombination in the observed trPL decay, we employed trSPV measurements to specifically focus on charge-separating phenomena at the CsPbI$_3$/TiO$_2$ interface.

In the trSPV experiments on CsPbI$_3$/TiO$_2$ systems as depicted in Fig. 3**C**, the signal is directly proportional to the charge extracted (Fig. 3**D**). The time-resolution allows for a differentiation of extracted charge carriers from other phenomena creating separated charges, such as (de-)trapping, recombination, or ionic displacement.[41][42] We show trSPV curves of mono-doped and co-doped samples in Fig. 3**D**. The initial slope in the nanosecond-regime of the transient curve is proportional to the extraction rate across the CsPbI$_3$/TiO$_2$ interface. Additionally, the absolute amplitude is directly correlated to the maximum charge separation possible within the bilayered system at the charge carrier density induced by the laser pulse. In the experiment, we controlled the laser fluence at 22.5 nJcm$^{-2}$ for all samples. The sample co-doped with 0.5% Nb(V) and 0.1% Sn(IV) exhibits the highest extraction rate as well as the largest charge separation across all the tested excitation wavelengths (**Fig. S10D**). The increased charge separation for co-doped TiO$_2$ corroborates the suggestions made in the trPL analysis. We can, therefore, assign the shorter lifetime $\tau_{fast}$ to an increased electron extraction rate instead of interface recombination.

Having derived faster and larger charge separation across the interface between $CsPbI_3$ and co-doped $TiO_2$ from transient measurements, the question of the partially quenched ssPL signal remains open. We suggest that in steady state, extraction rate and back-transfer rate do not necessarily equilibrate. Instead, extracted electrons can still recombine non-radiatively via interface defects with holes in the perovskite and are therefore removed from radiative recombination. The increased extraction rate of the co-doped $TiO_2$ can therefore lead to a reduction of the photoluminescence quantum yield, resulting in a quenched PL signal.[43]

Interestingly, the trSPV curves of doped $TiO_2$ in Fig. 3**D** exhibit an overshoot below zero at millisecond timescales. We hypothesize that the signal can be assigned to the capture of electrons in deep trap states. Due to their depth, de-trapping times are significantly longer than for shallow trap states. After full back-transfer of electrons from the ESC and recombination, the remaining captured electrons are then responsible for the SPV signal in the opposite direction. The overshoot of the signal is not influenced by our co-doping strategy, implying that the deep trap states are located on the exposed surface of the perovskite.

We have demonstrated that the charge extraction rates across the $TiO_2/CsPbI_3$ interface can be improved by Nb(V) and Sn(IV) co-doping. Since trSPV measurements mostly allow for qualitative rather than quantitative characterization, we developed a simulation approach based on DD to further extract crucial parameters from the trSPV curves.

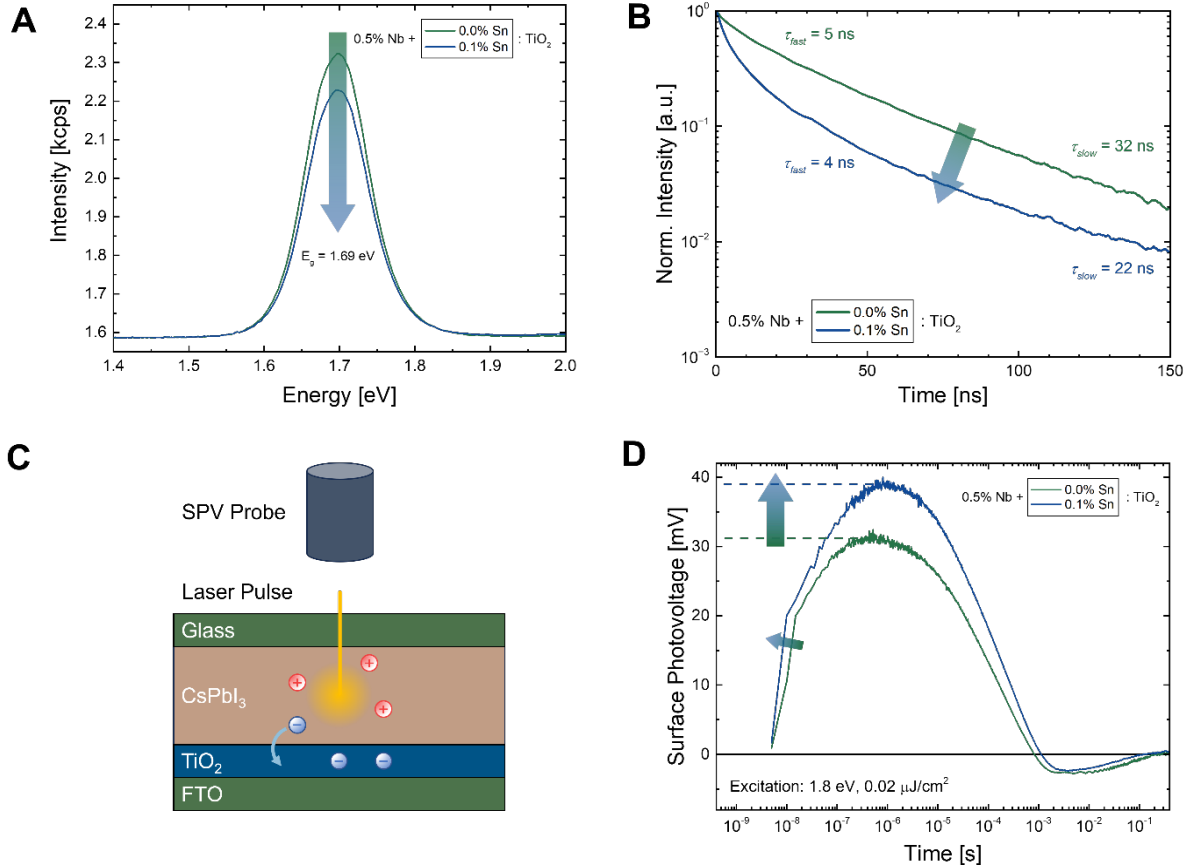

**Fig. 3. Co-doping of $TiO_2$ with Nb(V) and Sn(IV) reduces non-radiative recombination and improves the charge extraction efficiency. (A)** ssPL intensity decreases upon co-doping 0.5% Nb $TiO_2$ with Sn(IV). **(B)** trPL measurements showing faster PL decay at the Sn(IV)-free $TiO_2$/$CsPbI_3$ interface. **(C)** Illustration of the origin of the SPV signal. Upon excitation, electrons will be extracted from $CsPbI_3$ to (co-doped) $TiO_2$. The separated charge is detected as a positive signal by the SPV probe in a fixed capacitor arrangement. **(D)** The corresponding trSPV curve qualitatively shows that, in co-doped $TiO_2$, more electrons are extracted at a higher rate compared to mono-doped $TiO_2$.

**Simulation of trSPV Curves**

By using KPFM, PL and SPV, we have demonstrated that the charge extraction across the $TiO_2$/$CsPbI_3$ interface can be improved via Nb(V) and Sn(IV) co-doping, possibly due to an elevated Fermi level and associated increase in conductivity. To fully interpret the trSPV data in terms of carrier photogeneration, diffusion, and extraction, we employed 2D DD simulations. We report more details on the simulations in the SI, together with the applied parameters (Tab. S6 – **Tab. S9**).

In the DD simulation, we account for the development of the electric field according to Poisson's equation., as described by **Eq. 3** – **Eq. 8** in **Sec. S3.11** of the SI. Both the trSPV curves of 0.5% Nb(V) doped and 0.5% Nb(V) and 0.1% Sn(IV) co-doped $TiO_2$ were fitted with the DD model, which features an non-doped perovskite layer, an non-doped but defective $TiO_2$ layer, an interface layer between the two, and a Schottky-type contact at the back of $TiO_2$. The two transients will be called *mono-doped* and *co-doped* in the following.

While certain parameters were common to the mono- and co-doped fit, such as the perovskite carrier mobility ($\mu_{n,p}$) and the non-radiative characteristic time in $CsPbI_3$ (). Other parameters, such as the non-radiative recombination velocity at the $CsPbI_3/TiO_2$ interface ($\nu_{interf,e}$ and $\nu_{interf,h}$), and the donor trap parameters in $TiO_2$, were fitted independently on each curve.

For both configurations, the band alignment between $CsPbI_3$ and $TiO_2$ is governed by the affinity $\chi_{pero}$, while the bulk diffusion length depends on mobility and lifetime In addition to recombination in the bulk, we expect non-radiative recombination at the $CsPbI_3/TiO_2$ interface, which depends on the $TiO_2$ composition since previous works have pointed out the presence of defects on the $TiO_2$ surface such as oxygen vacancies and extrinsic adsorbed species.[16][30-33]

To model the complex doping picture of $TiO_2$, we chose to add single-energy level ($E_{tD}$) bulk donor traps with defined density $N_{tD}$ and electron capture cross section $\sigma_{etD}$, and we let those parameters vary in order to fit the trSPV curves. These traps modify the carrier concentration in $TiO_2$, inducing an n-type doping nature as well as storing photoelectrons injected in $TiO_2$ from $CsPbI_3$.

In Fig. 4A, we report the result of the fit on the mono-doped $TiO_2$ trSPV curve employing the parameters listed in **Tab. S6**. On the experimental curve, we can identify four different regions: a rapid rise of SPV signal at $t < 10^{-8}$ s, followed by a slower rise between $10^{-8}$ s $< t < 10^{-6}$ s, a decay of SPV signal between $10^{-6}$ s $< t < 10^{-3}$ s, and a negative tail after $t > 10^{-3}$ s. The positive polarity is the result of an accumulation of positive charges close to the surface, while negative charges are driven towards the bulk. This is in line with the result of our simulations, where the positive sign is due to electron diffusion towards the CsPbI$_3$/TiO$_2$ interface and subsequent injection into the TiO$_2$ layer, while holes only diffuse towards the buried interface and cannot be injected into the TiO$_2$ layer due to the energetic offset. Being governed by diffusion, we can estimate the transit time of electrons across the CsPbI$_3$ layer to 33 ns via , where  is the CsPbI$_3$ thickness, and $D$ is the diffusion coefficient ($D = kT\mu_{e,h}/q$, with $kT$ being the electron thermal energy, $\mu_{e,h}$ and $q$ being the electron mobility and charge, respectively).  corresponds to the simulated SPV rise time in Fig. 4A. Electron injection in TiO$_2$ is allowed by the CBM alignment between CsPbI$_3$ and TiO$_2$, and is further favored by the large concentration of deep donor traps in the TiO$_2$ layer. In fact, these traps create states within the TiO$_2$ bandgap, which capture and store electrons quickly, allowing further injection of electrons from CsPbI$_3$.

While the fast SPV rise and decay are decently reproduced by our model, the current set of parameters does not yet allow for modification of the simulated curve shape. To improve the fit, we introduced a delay component in the SPV rise, which allowed the SPV to continue growing slowly after the fast rise. This was accomplished by adding shallow electron acceptor traps close to the CBM of CsPbI$_3$ (s. orange curve in Fig. 4A, best-fit parameters in **Tab. S8**). With the newly found CsPbI$_3$ carrier mobility, we find the electron diffusion time across the CsPbI$_3$ layer, which coincides with the kink in the experimental trSPV curve at  = 12 ns. Simultaneously, shallow traps in CsPbI$_3$ capture electrons with an average capture time of $1/(\sigma_{etA} \cdot v_{th} \cdot N_{tA}) = 10$ ns, where $N_{tA}$ is the acceptor trap density in CsPbI$_3$, $\sigma_{etA}$ is the electron capture cross section associated with the traps, and $v_{th}$ is the electron thermal velocity.

Between $10^{-8}$ s $< t < 10^{-6}$ s, the signal rise is no longer dominated by electron diffusion, but rather by electron de-trapping from the shallow traps to the conduction band and subsequent diffusion towards TiO$_2$. The electron re-emission time can be estimated at  $= \sigma_{etA} \cdot v_{th} \cdot N_C \cdot e^{(E_{tA}-E_{CBM,pero})/kT})^{-1} = 36$ ns, which is close to the SPV peak time.

Here, $N_C$ is the conduction band effective density of states, ($E_{tA} - E_{CBM,pero}$) is the trap energy with respect to the CBM, and $kT$ is the electron thermal energy.

With the same principle, we can estimate the electron capture time of TiO$_2$ traps by employing the trap density $N_{tD}$ and the electron capture cross section $\sigma_{etD}$. We find $= 1/(\sigma_{etD} \cdot v_{th} \cdot N_{tD}) = 2.7 \cdot 10^{-15}$ s, implying that it does not limit electron diffusion towards TiO$_2$. In addition, we find $= 2.3 \cdot 10^{-3}$ s for the electron release time from TiO$_2$ traps. roughly corresponds to the SPV decay time in our model, since the SPV decay is governed by electron re-emission from TiO$_2$ traps, which is followed by recombination across the interface, with holes present in the CsPbI$_3$ layer. Having obtained a satisfying fit, we can now evaluate the impact of each parameter on the simulated curve.

We found that the SPV signal height depends mostly on generation and recombination-related parameters, which control the number of electrons that can be injected in TiO$_2$. In particular, it depends on the laser fluence $F_{laser}$, on the bulk non-radiative characteristic time in CsPbI$_3$ () and carrier mobility in CsPbI$_3$ ($\mu$), which control the bulk diffusion length, and $v_{interf,e}$ and $v_{interf,h}$. The signal height also depends on the shallow electron acceptor trap parameters, since these defects are capable of recombining electrons with holes in bulk CsPbI$_3$, in addition to storing electrons.

Crucially, $v_{interf,e}$ and $v_{interf,h}$ should strongly depend on the TiO$_2$ composition. While in our model, $v_{interf,e}$ controls both the SPV amplitude and decay time, $v_{interf,h}$ influences primarily the SPV amplitude. If we compare the trSPV curves of mono- versus co-doped TiO$_2$, $v_{interf,h}$ decreases from 17.0 cm·s$^{-1}$ to 0.098 cm·s$^{-1}$, with the other parameters being set according to **Tab. S8**. In addition, by increasing the TiO$_2$ donor trap energy level from 2.673 eV to 2.758 eV from the TiO$_2$ valence band maximum (VBM), the fit further improves. The obtained parameter values closely match the best-fit parameters for co-doped TiO$_2$ found in **Tab. S9**. The best-fit curves for both mono- and co-doped TiO2 are shown in Fig. 4**B**. Therefore, according to our model, the signal height difference between the two curves is mainly due to an improvement of the CsPbI$_3$/TiO$_2$ interface, which translates into a reduced $v_{interf,h}$ of two orders of magnitude in co-doped TiO$_2$ with respect to mono-doped TiO$_2$. **Fig. S17** reports the signal height dependence on $v_{interf,h}$.

The other crucial difference between the mono- and co-doped parameters is the TiO$_2$ donor trap energy level. Since these traps control the TiO$_2$ equilibrium Fermi level ($E_F$), a *WF* difference of about 80 meV is found ($\phi_{mono} - \phi_{co} = 80$ meV), implying that the co-doped TiO$_2$ $E_F$ moves closer to the CBM compared to mono-doped TiO$_2$, synonymous with larger n-type character.

To further consolidate the improvement in the extraction properties, Fig. 4**C** shows the total extracted electron concentration for each of the two fitted configurations. In Fig. 4**D**, we report the band diagram out of equilibrium for $t = 0.35$ μs, i.e. the SPV peak, for the fitted co-doped TiO$_2$. The lack of doping in the CsPbI$_3$ layer manifests in the absence of electric field across the layer, where the equilibrium Fermi level is controlled by TiO$_2$. If a doped HSC was deposited on CsPbI$_3$, a constant electric field would appear across the layer, improving extraction with respect to a purely diffusive scenario. In mono- or co-doped TiO$_2$ we notice a potential drop, synonym of a mismatch between the contact WF ($\phi_{cont} = 4.35$ eV), and the TiO$_2$ WF at the top interface ($\phi_{co} = 4.003$ eV, $\phi_{mono} = 4.085$ eV). The large potential drop in TiO$_2$ is due to the large donor trap concentration, which partially screens the back contact.

Page 17 of 27

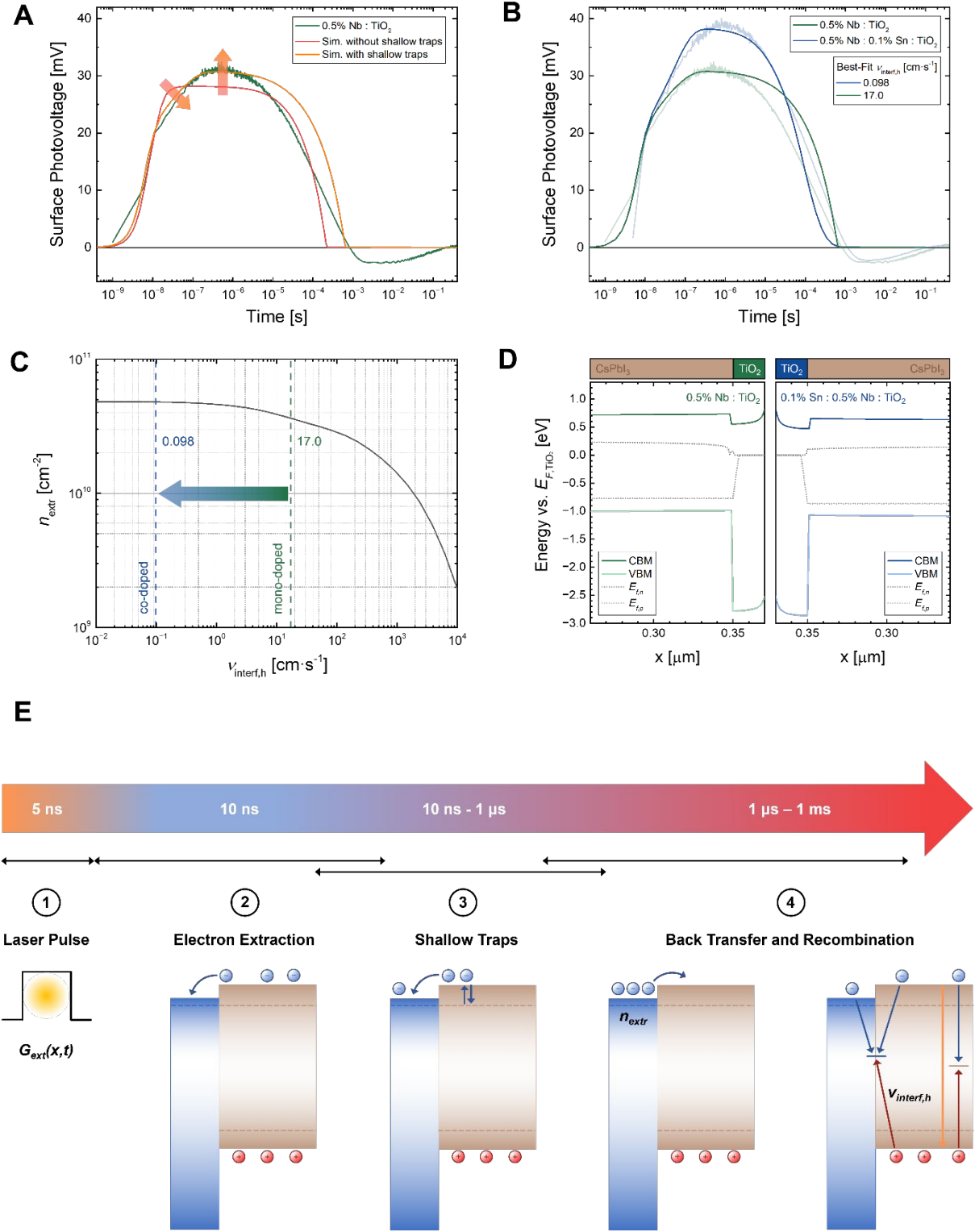

Fig. 4. 2D DD simulations of recorded trSPV data reveal a reduction of the interface hole recombination velocity by two orders of magnitude. (A) The match between the original data (green) and the simulated curve (red) improves after consideration of shallow electron traps in the CsPbI$_3$ layer (orange). (B) In the simulation model, the signal height is sensitive to the interface hole recombination velocity $v_{interf,h}$. Fitting the original data yields 17.0 cms$^{-1}$ for mono-doped TiO$_2$ and 0.098 cms$^{-1}$ for co-doped TiO$_2$. (C) Reduction of $v_{interf,h}$ leads to a higher concentration of extracted electrons. (D) Calculated out-of-equilibrium band diagram at maximum charge separation (SPV peak at $t = 0.35\mu s$). (E) DD simulations enabled us to associate decelerated electron extraction (10 ns – 1 µs) with the presence of shallow traps in CsPbI$_3$. For simplicity, the schematic disregards band-bending effects. (1) At $t = 0$, electrons (blue spheres) and holes (red spheres) are photo-generated in CsPbI$_3$ (brown box). (2)

Electron extraction to the TiO$_2$ ESC (blue box) is the reason for the charge separation at around 10 ns as measured with trSPV. (3) The kink in the trSPV curve between 10 ns – 1 µs implies ongoing charge separation by extraction at decelerated velocity until a maximum of extracted electrons $n_{extr}$ is reached. The kink was reproduced by DD simulation via introduction of shallow traps in CsPbI$_3$. (4) Signal depletion between 1 µs – 1 ms is caused by charge carrier recombination via different pathways, comprising electron back-transfer and subsequent classical radiative or non-radiative pathways, or interface recombination without back-transfer.

**Photovoltaic Performance**

We tested how doping affects the WF of the ESC and its consequences on the charge carrier dynamics when contacted with CsPbI$_3$. Finally, we implemented the developed (co-doped) TiO$_2$ layer in full photovoltaic devices in n-i-p architecture as shown in **Fig. 5A**, with spiro-OMeTAD as HSC and gold as the back contact. The resulting J-V characteristic curves show, on average, a slightly enhanced $V_{OC}$ by 12 meV from 1.167 V to 1.179 V (**Fig. S12**). Most evident, however, is the FF increase by 2.0% from 78.8% to 80.8%, as shown statistically in Fig. **5C**. These factors lead to a generally improved PCE by 1% from 16.4% to 17.4% (Fig. **5C**), with champion pixels of 17.2% for the mono-doped TiO$_2$ and 18.0% for the co-doped TiO$_2$ (Fig. **5B**).

Further, we assessed the stability of the PSCs via maximum power point (MPP)-tracking under 1-sun continuous illumination according to the ISOS-L1I protocol.[44][45] Fig. **5D**, for comparability, shows the normalized performance of our solar cells. Absolute performances are referred to in **Fig. S14**. All solar cells show an initial fast decay of efficiency within the first 30 h, the so-called burn-in.[46] This burn-in phase is typical for device structures containing spiro-OMeTAD, and was associated with Li$^+$ ion migration.[47] The end of the burn-in phase is marked by the time $T_S$ when the curve transitions into a linear regime which is set to 50 h for all traces in Fig. **5D**. It is evident that, in the average of around 10 pixels, co-doped TiO$_2$ at 0.5% Nb(V) and 0.1% Sn(IV) outperforms the Nb(V) mono-doped champion as well as the sample over-doped with 0.5% Nb(V) and 2.0% Sn(IV). Fitting the traces with linear regression starting at $T_S = 50$ h, as indicated in Fig. **5D**, yields degradation rates of –0.06·10$^{-4}$ %/h, –1.50·10$^{-4}$ %/h, and –5.24·10$^{-4}$ %/h, respectively. From those slopes in the linear regime, we calculated time at which the cells have degraded to 80% of the initial stabilized PCE ($T_{S80}$),[33] yielding projected $T_{S80}$ lifetimes of around 25'000 h, 970 h and 280 h (**Fig. S16**). The linear degradation observed in our over-doped $TiO_2$ system suggests the continuous formation of deep, detrimental defects. In case of the ESC co-doped at optimum, the formation of those defects is suppressed. Based on *ab initio* density functional theory calculations, compound defect $[2Cs]_{Pb}$ has been suggested as the only electronically active defect introducing deep trap levels under operating conditions in $CsPbI_3$.[48] We speculate that, by elevating the Fermi level in $TiO_2$, the resulting change of chemical potentials inhibits the formation of a detrimental defect type. The improved stability is, therefore, a result of the optimized extractive and recombinative properties of the co-doped $TiO_2$ ESC.

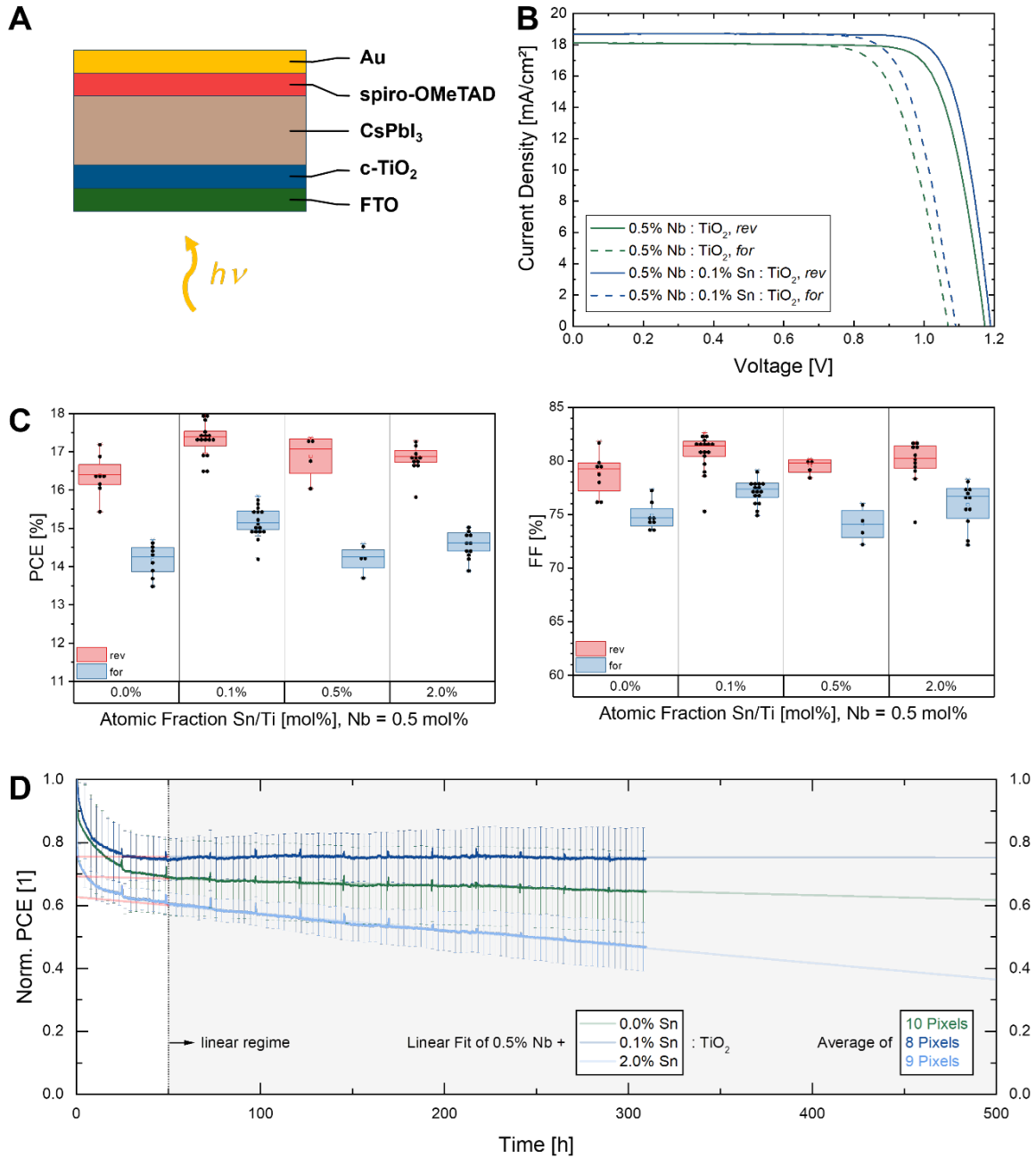

**Fig. 5. The co-doping approach improves the photovoltaic parameters.** (**A**) General device structure of the fabricated solar cells with Nb(V) and Sn(IV) being present in c-TiO$_2$. (**B**) *J-V*-Characteristic of the best performing devices, showing increased $V_{OC}$ and *FF* for the co-doped sample. (**C**) Statistical evaluation of the solar cells' photovoltaic parameters shows significant improvement in PCE from 16.4% to 17.4%, mainly originating from *FF* improvement from 79% to 82%. (**D**) Normalized device performance tested in MPP-tracking under continuous AM1.5G illumination. The co-doped champion does not exhibit significant performance loss over 300 h after the initial burn-in phase. $T_{S80}$ projections were obtained via linear regression of the MPP-tracking data.

## Conclusion

The aim of the presented study was the development of a strategy to improve the stability of CsPbI$_3$ PSCs by tailoring the Fermi-level position via (co-)doping of the TiO$_2$ ESC. Via compositional control of TiO$_2$, we improved electron extraction from CsPbI$_3$. Qualitative assessment of trSPV curves of the bilayered system revealed an optimum extraction rate at co-doping levels of 0.5% Nb(V) and 0.1% Sn(IV) in TiO$_2$. Further, we introduced a method to fit trSPV curves with a DD model of the system. The model enabled us to extract fundamental parameters, such as defect densities, interface recombination velocities, and extracted charge concentrations. We found that implementing co-doped TiO$_2$ leads to reduced interface hole recombination velocities and, consequently, increased concentration of extracted electrons. When applied in full photovoltaic devices, the co-doped ESC outperforms the mono-doped ESC in all relevant figures of merit. Consequently, the addition of 0.1% Sn(IV) increased the projected $T_{S80}$ lifetimes by a factor of 25. These results show that a small compositional intervention at the interface has the potential to translate to strong long-term stabilization of the photovoltaic device.

## Acknowledgments

The author wants to thank Markus Johannes Beckedahl for support in stability measurements, as well as the other technicians Carola Ferber and Hagen Heinz for making the laboratory run smoothly. Furthermore, thanks to Pascal Rohrbeck and Yenal Yalçınkaya for support in KPFM measurements as well as to Chiara Frasca, Florian Scheler, Manuel Felipe Vasquez Montoya, and Maxim Simmonds for valuable discussions. We gratefully acknowledge Dr. Thomas Dittrich for providing the HZB SPV laboratory facilities. This project has received funding from the European Union's Framework Program for Research and Innovation HORIZON EUROPE (2021-2027) under the Marie Skłodowska-Curie Action Postdoctoral Fellowships (European

Fellowship) 101061809 HyPerGreen. We acknowledge HyPerCells – a joint graduate school of the University of Potsdam and the Helmholtz-Zentrum Berlin – and the Deutsche Forschungsgemeinschaft (DFG, German Research Foundation), project number 423749265, SPP 2196 (SURPRISE) for funding. M. S. acknowledges the Heisenberg program of Deutsche Forschungsgemeinschaft (DFG, German Research Foundation) for funding – project number 498155101. M. S. further acknowledges the CUHK Vice-Chancellor Early Career Professorship for funding. We thank HZB for the allocation of synchrotron radiation beamtime. This work was supported by the French government in the frame of the program of investments for the future (Programme d'Investissement d'Avenir ANR-IEED-002-01).

## Author Contributions

T. W. G. conducted the research and fabricated all solar cells and samples, recorded, and analyzed EQE, ssPL and UV/Vis spectra and *J-V*-scans. KPFM images were taken and evaluated by T. W. G. with the equipment of and training by S. A. W.. D. R., A. M. and J.-P. K. developed the DD model. D. R. fitted the trSPV data using facilities of J.-B. P., and P. S.. A. M. performed trSPV measurements and provided fundamental discussion and guidance. H. K. and N. T. P. H. recorded and evaluated LT-MPP data. R. F., E. H. and A. S. conducted the HAXPES measurements. R. F. analyzed the HAXPES data, with close support from R. G. W. and M. B. in the interpretation of the results. E. G. P. measured and analyzed trPL with analytical support by M. S.. G. V. S., Z. I., and Z. L. N. provided fruitful discussions. T. W. G. prepared the manuscript revisited by A. M., E. U., and A. A. Q. W., A. M. and A. A. took the supervision of the project.

## Conflict of Interest

The authors do not declare conflict of interest.